# Ultrafast Li Electrolytes Based on Abundant Elements: $Li_{10}SnP_2S_{12}$ and $Li_{11}Si_2PS_{12}$

A. Kuhn [a], Oliver Gerbig [a], Changbao Zhu [a], Frank Falkenberg [a], Joachim Maier [a], and Bettina V. Lotsch [a,b,c]

**We report on the synthesis and characterization of two solid electrolytes, $Li_{10}SnP_2S_{12}$ and $Li_{11}Si_2PS_{12}$, which are based exclusively on abundant elements. Both compounds feature extremely high Li ion diffusivities, with the Si-based material even surpassing the present record holder, the related electrolyte $Li_{10}GeP_2S_{12}$. The structure and dynamics were studied with multiple complementary techniques and the macroscopic diffusion could be traced back to fast Li ion hopping in the crystalline lattice.**

In 2011, the new solid lithium electrolyte $Li_{10}GeP_2S_{12}$ (LGPS) was reported, featuring liquid-like Li ion conduction in a crystalline solid matrix.[1,1a] The ultrafast room temperature transport of tetragonal LGPS with a conductivity of several mS/cm came as a surprise as it exceeds the values of the best crystalline Li conductors by one order of magnitude. Beside the fundamental aspect of unexpectedly high Li mobility the authors reported a large effective electrochemical window rendering this material interesting for Li based batteries. The high cost of germanium hampers such application. Therefore, there has been a strong upsurge of interest recently in synthesizing LGPS-type materials based on the homologous elements Si and Sn. A theoretical study published by Ceder and coworkers highlights the potential of such hypothetical tetragonal LGPS-type Li ion conductors.[2,3] It was predicted that the beneficial properties of tetragonal LGPS are retained when Ge is replaced by Si or Sn.

Here, we report the synthesis of Ge-free LGPS-type electrolytes, namely tetragonal $Li_{10}SnP_2S_{12}$ (LSnPS) and $Li_{11}Si_2PS_{12}$ (LSiPS).[3] Both show extremely high $Li^+$ diffusivities with the values of the Si compound exceeding those of LGPS, the present record holder. The materials were comprehensively characterized both with respect to their structure and Li ion dynamics. These results are compared with those recently obtained for $Li_{10}GeP_2S_{12}$ and $Li_7GePS_8$, two members of the tetragonal LGPS solid solution which both showed very similar Li diffusivities.[5] In perfect agreement with the theoretical predictions in Ref. [2], LSiPS shows an even higher Li diffusivity than LGPS while LSnPS has a slightly lower Li diffusivity. We further demonstrate that the synthesis of tetragonal LiSiPS is only possible at high pressure, hence limiting scalability, while the synthesis of LSnPS can easily be scaled up to larger volumes.

Tetragonal $Li_{10}SnP_2S_{12}$ was prepared from elemental Sn, P, S, and $Li_2S$. The dry starting materials were first mechanically treated in a ball mill under argon for 2 days. Pellets were pressed from the obtained amorphous precursor and heated in an evacuated quartz ampoule according to the following temperature program: 30 Kh$^{-1}$ → 613 K (10h) → 30 Kh$^{-1}$ → 653 K (10h) → 30 Kh$^{-1}$ → 723 K (2d). In the first step at 613 K, the material crystallizes in the orthorhombic modification (cf. $Li_4SnS_4$[8]) which then transforms to the desired tetragonal structure at 653 K. The subsequent sintering step at 723 K improves the crystallinity of the sample. The tendency of the larger Sn$^{IV}$ to reside in six-fold rather than four-fold coordination as desired for LSnPS is reflected by the presence of side phases with edge-sharing $SnS_6$ building units (such as $Li_2SnS_3$) when stoichiometric amounts of the starting materials are used. A 10-20% excess of $Li_2S$, however, completely prevents the formation of these layered side phases. It should be noted that we used a slight excess of S yielding approx. 1 atm S at the reaction conditions in order to ensure complete oxidation of Sn and P.

The preparation of phase-pure tetragonal $Li_{10}SiP_2S_{12}$ was not successful by means of conventional solid-state synthesis, although the desired product appeared to be present as a low-percent (< 10 %) side phase for samples prepared from the melt. The main phase at all temperatures between 573 K and 1023 K was the orthorhombic modification of the LiSiPS solid solution, which was reported by Kanno et al. in 2002.[5] However, since for both LGPS and LSnPS the tetragonal modification has a slightly higher density than the orthorhombic one (for LGPS, compare Ref [1,3,5] and Ref. [9]), we assumed that tetragonal $Li_{10}SiP_2S_{12}$ should be accessible via high-pressure synthesis. Indeed, when orthorhombic $Li_{10}SiP_2S_{12}$ was subjected to pressures of 5GPa at 823 K, the tetragonal structure was partially formed, next to a side phase. A SEM-EDX analysis of the product revealed that the Si:P ratio was no longer homogeneous after the high-pressure treatment. Some crystallites were enriched with Si (Si:P ≈ 2:1), while others were depleted of Si. From considerations of the ionic radius of Si (see discussion below), we assumed that the crystallites with Si:P ≈ 2:1 represented the tetragonal phase. Indeed, when a material with the target stoichiometry $Li_{11}Si_2PS_{12}$ and orthorhombic symmetry was separately prepared in a conventional solid-state synthesis and used as a precursor for the high-pressure synthesis, phase-pure tetragonal LSiPS was obtained. High-pressure treatment at 723 K with pressures in the range $3 < p < 5$ GPa furnished the best results. The precursor for the high-pressure synthesis, i.e. orthorhombic $Li_{11}Si_2PS_{12}$, was again prepared from $Li_2S$, Si, P, and a slight excess of S. 1 g of the mixture was mechanically treated in a high-energy ball mill for 2 days, followed by heat treatment in an evacuated quartz tube at 823 K for 4 days.

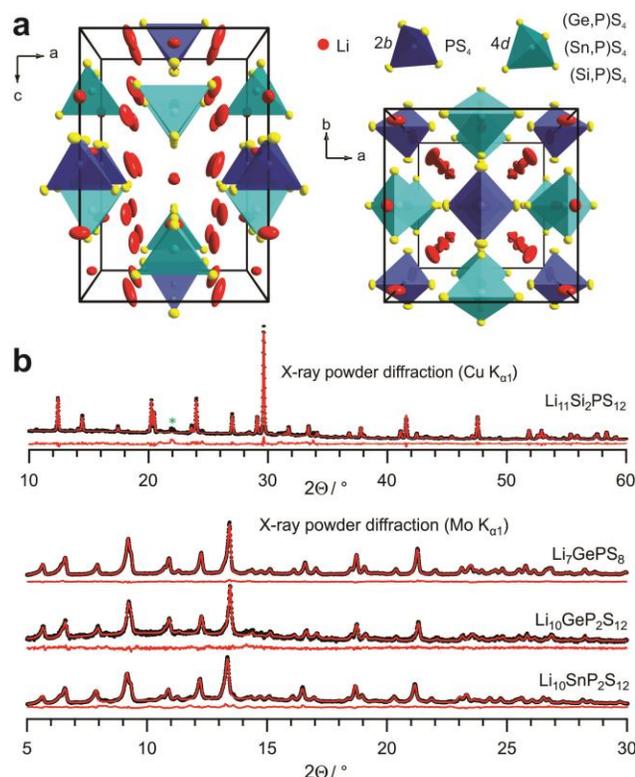

**Figure 1: a:** Crystal structure of tetragonal LGPS as obtained from single-crystal X-ray diffraction.[6] **b:** X-ray powder diffraction and Rietveld refinement of $Li_{11}Si_2PS_{12}$ and $Li_{10}SnP_2S_{12}$ in comparison to previously reported $Li_{10}GeP_2S_{12}$ and $Li_7GePS_8$.[4] The side phase is marked by a green asterisk.

Fig. 1a depicts the unit cell of the generic tetragonal LGPS structure as obtained from single-crystal diffraction of $Li_{10}GeP_2S_{12}$.[6] The structure contains $PS_4$ and $GeS_4$ tetrahedra which are charge-compensated by Li ions. In the tetragonal LGPS structure type, there exist two sets of tetrahedra with the central atom on a 4$d$ and 2$b$ site, respectively, with the tetrahedra around the 4$d$ site being considerably larger. Consequently, in tetragonal LGPS, the 4$d$ site is occupied by both Ge and P, while the smaller 2$b$ site is solely occupied by P.[1,3,5] The occupancy of the mixed occupied 4$d$ site ($Ge_xP_{1-x}$) can be varied in the range of $0.5 \leq x(Ge) \leq 0.75$ with the end members of the accessible range of the solid solution being roughly $Li_7GePS_8$, and $Li_{10}GeP_2S_{12}$.[3,6] Fig. 1b shows the XRD patterns with single-phase Rietveld refinements for $Li_{10}SnP_2S_{12}$ and $Li_{11}Si_2PS_{12}$ in comparison with the patterns for tetragonal LGPS ($Li_7GePS_8$ and $Li_{10}GeP_2S_{12}$). All samples show the desired tetragonal structure[1,5] and are phase-pure on the XRD level except for a weak additional reflection for LSiPS (marked by an asterisk and tentatively ascribed to a high-pressure modification of S[10]). The cell parameters are listed in Table S1. According to the Rietveld refinement (see Supporting Information S2), in $Li_{10}SnP_2S_{12}$ the 4$d$ site is occupied by Sn and P ($x(Sn) = 0.47$), whereas again, the 2$b$ site is occupied by P only. This is in line with the results from single-crystal X-ray diffraction recently published by Bron *et al.*[3b] The accessible range of the LSnPS solid solution is much narrower than in case of LGPS – we obtained the tetragonal modification only for values of $x(Sn)$ very close to 0.5. For other Sn/P ratios, two phases were obtained: tetragonal LSnPS of the composition $Li_{10}SnP_2S_{12}$ and the orthorhombic modification as side phase. This can be explained by the larger ionic radius of $Sn^{IV}$ as compared to $Ge^{IV}$, rendering a higher occupancy of this position relative to $P^V$ energetically unfavourable. The opposite is true for LSiPS: $Si^{IV}$ is only slightly larger than the isoelectronic $P^V$ ion. Therefore, in order to stabilize the tetragonal modification with two sets of differently-sized tetrahedra, the $4d$ site has to be occupied by Si to a much higher extent as compared to Ge or Sn. This explains why the tetragonal modification is obtained for the stoichiometry $Li_{11}Si_2PS_{12}$ rather than $Li_{10}SiP_2S_{12}$.

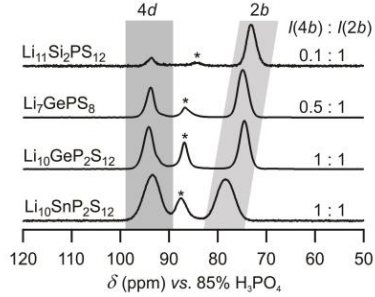

**Figure 2:** $^{31}$P MAS NMR spectra ($\nu_{rot}$ = 12 kHz, $B_0$ = 9.4 T) of the different isostructural materials crystallizing in the tetragonal LGPS-type. The lines for $Li_7GePS_8$ and $Li_{10}GeP_2S_{12}$ as well as the assignment of the lines were taken from Ref. [4]. The asterisks denote impurities with a chemical shift typical of the orthorhombic modification.

$^{31}$P MAS NMR was used in order to probe the relative amount of P residing on the $4d$ and $2b$ sites (*cf.* Fig. 1a). This is of special importance for the structural elucidation of the LSiPS sample because here, this information is not accessible from X-ray diffraction since $Si^{IV}$ and $P^V$ are isoelectronic. Fig. 2 shows the $^{31}$P MAS NMR spectra of the tetragonal LGPS-type electrolytes. The spectrum of $Li_{10}SnP_2S_{12}$ is very similar to that of $Li_{10}GeP_2S_{12}$: As expected from Rietveld refinement, the relative intensity of the signals assigned to the $4d$ and $2b$ sites is 1:1 within 5% error. The third signal shows the chemical shift typical of the orthorhombic modification, which is, however, not observed in the X-ray patterns. Therefore, we assign it to the presence of a side phase of low crystallinity, which resembles the local structure of the orthorhombic modification. For $Li_{11}Si_2PS_{12}$, the structural model assumed above – judging only from the stoichiometry of the sample – is largely verified. Only a small amount of P resides on the mixed-occupied $4d$ site (~10%). Thus, the occupancy of the $4d$ site for all LGPS-type samples follows a clear trend. For the largest ion, Sn, the occupancy is close to $x(Sn) \approx 0.50$. For Ge, the occupancy ranges from $0.5 < x(Ge) < 0.75$, while for the small Si, $x(Si) \approx 0.95$.

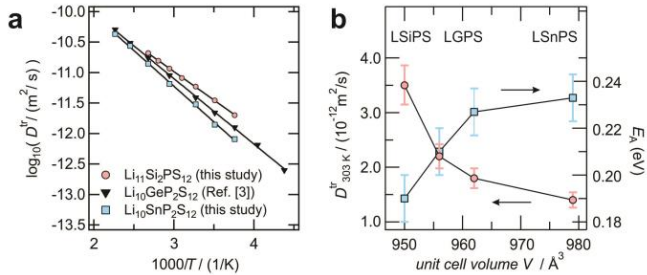

**Figure 3: a:** Li tracer-diffusion coefficients of LGPS-type electrolytes as measured by $^7$Li PFG NMR. **b:** correlation between diffusion parameters and unit cell volume for LGPS-type electrolytes.

In order to characterize the Li diffusivity in the new LGPS-type materials $Li_{10}SnP_2S_{12}$ and $Li_{11}Si_2PS_{12}$, we performed $^7$Li PFG NMR measurements. The obtained Li tracer-diffusion coefficients $D^{tr}$ (3D diffusion, *cf.* Ref. [4]) are shown in Fig. 3a in comparison with those previously reported for $Li_{10}GeP_2S_{12}$.[4] Note that in all cases, the measured diffusion coefficients can clearly be assigned to diffusion in the bulk of the tetragonal LGPS-type electrolytes, since the quadrupolar structure of the decaying NMR signal showed the finger-print (see Fig. 4a and 4b) of the tetragonal modification, which is distinct from that of orthorhombic or amorphous side phases. The diffusivity of $Li_{10}SnP_2S_{12}$ is slightly lower than that of $Li_{10}GeP_2S_{12}$ and the activation energy is slightly higher (0.23(1) eV vs. 0.21(1) eV). In contrast, the diffusivity of $Li_{11}Si_2PS_{12}$ is even higher than that of $Li_{10}GeP_2S_{12}$ with a slightly lower activation energy (0.19(1) eV vs. 21(1) eV). We would like to point out that this trend is – both qualitatively and quantitatively – in very good agreement with theoretical calculations by Ong et al.[2] as presented in

Table 1. Note that the theoretical diffusion coefficients taken from the MD simulation in Ref. [2] have been extrapolated from 600 K down to room temperature in order to compare them with our experimental data. As shown in Fig. 3b, a clear correlation between the unit cell volume of the tetragonal LGPS-type materials and their diffusion parameters is observed. LSiPS shows the smallest unit cell volume and enhanced Li diffusivity, while LSnPS shows the largest unit cell volume and the lowest diffusivity.

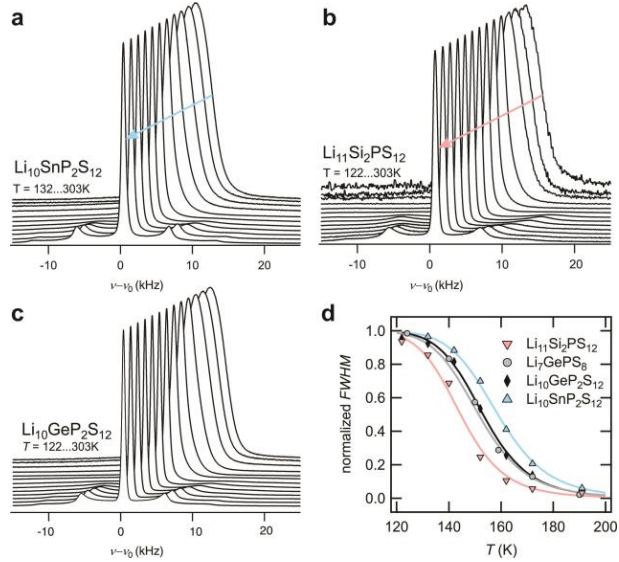

**Figure 4:** Temperature dependent $^7$Li NMR spectra of (**a**) $Li_{10}SnP_2S_{12}$, (**b**) $Li_{11}Si_2PS_{12}$, and (**c**) $Li_{10}GeP_2S_{12}$ (taken from Ref. [4]). **d:** comparison of the normalized narrowing curves of the LGPS-type electrolytes.

While $^7$Li PFG NMR probes Li ion dynamics occurring on the micron scale and in the time scale of milliseconds, NMR relaxometry is sensitive to Li site-to-site hopping on the Ångström scale and in the time scale of microseconds (transversal relaxation) or nanoseconds (longitudinal relaxation).[11] Thus, NMR relaxometry allows connecting the observed diffusion processes with their microscopic origin, i.e. site-to-site hopping of Li ions in the structure. Here, we measured the transversal relaxation – its Fourier transform being the static $^7$Li NMR spectrum – in order to probe the Li hopping processes. The temperature-dependent $^7$Li NMR spectra of LSnPS and LSiPS are displayed in Figs. 4A and 4B, respectively. As expected for isostructural compounds, the overall behavior of the line shape is very similar to that observed for LGPS[4] (see Fig. 4C). At low temperatures, the central transition of the $^7$Li line is broad and Gaussian-shaped representing the static homonuclear and heteronuclear dipolar interaction of the $^7$Li spins with their environment. At higher temperatures, as the Li jump rate exceeds the time scale of the dipolar interactions determined by the rigid-lattice line width, the line narrows successively showing a sharp Lorentzian-shaped line at higher temperatures. From the onset of the motional narrowing at $T_{MN}$, a jump rate can be assessed according to the narrowing condition $\tau^{-1} \approx \sqrt{M_{2\,\text{rigid lattice}}}$ whereby $M_{2\,\text{rigid lattice}}$ is the second moment of the rigid lattice line.[11] In Fig. 4D, the narrowing curves of the four LGPS-type electrolytes are compared. The trend already observed in PFG NMR is clearly reproduced. The Li jump rates at $T_{MN}$ amount to $1.5 \times 10^4$ s$^{-1}$ @ 125 K for LSiPS, $1.4 \times 10^4$ s$^{-1}$ @ 135 K for LGPS, and $1.4 \times 10^4$ s$^{-1}$ @ 145 K for LSnPS, respectively (values included in Fig. 5C, see Supporting Information for further details). As already discussed for the LGPS samples in Ref. [4], the distinct quadrupolar powder pattern, which is observed for both LSiPS and LSnPS at higher temperatures, represents the quadrupolar coupling between the quadrupolar moment of the $^7$Li spins and a mean electric field gradient they are exposed to while diffusing through the tetragonal lattice. As observed in the experiment, this mean electric field gradient retains the axial symmetry of the tetragonal structure ($\delta_Q = 23.5 \ldots 25.9$ kHz, $\eta_Q = 0$). Thus, the appearance of this distinct quadrupolar powder pattern indicates that the fast Li dynamics measured in fact occurs within crystallites of tetragonal LSnPS or LSiPS.

Impedance spectroscopy was applied in order to study the ionic conductivity of $Li_{10}SnP_2S_{12}$. Fig. 5A shows the temperature-dependent total conductivity and the bulk conductivity extracted from the impedance spectroscopy measurements for $Li_{10}SnP_2S_{12}$ (see Supporting Information for details). The bulk conductivity is activated with 0.25 eV, the room-temperature conductivity amounts to 4 mS/cm, again in remarkably good agreement with the values predicted from MD simulations[2] (0.24 eV and 6 mS/cm). Table S3 (see Supporting Information) summarizes the bulk conductivity data for LGPS-type electrolytes as obtained in our measurements. In order to determine the electronic contribution to the total conductivity, a dc polarization measurement was carried out at 473 K by applying a small current of 1 nA to a symmetric cell with ion blocking Au electrodes Au|LSnPS|Au. Fig. 5B shows the polarization curve. For a good solid electrolyte with vanishingly small electronic contribution one not only expects a considerable polarization but also a small chemical diffusion coefficient. Accordingly, a steady state could not be attained within reasonable waiting time, but an upper limit of the electronic conductivity corresponding to an electronic transference number of $t_{EON} < 1$ PPM could be safely determined from the absolute voltage values as well as from the time behaviour (cf. Supporting Information S5). Thus, LSnPS can be considered as a purely ionic conductor. The impedance of the pellet prior to and after the dc polarization measurement was equal within 1 %. For LSiPS, owing to the instability at sintering temperatures, the preparation of phase pure dense ceramic samples suitable for precise impedance spectroscopic experiments failed and we hence concentrate on the NMR results.

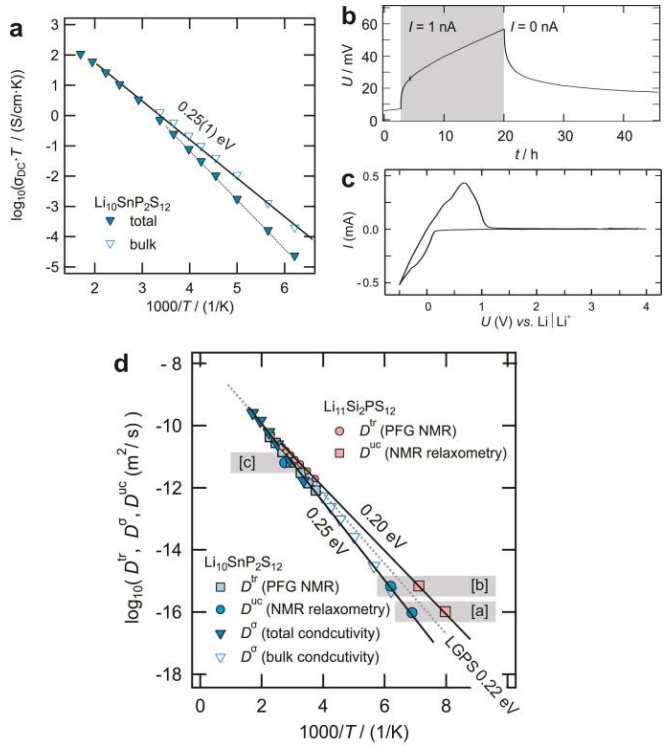

**Figure 5: a:** Total and bulk conductivity of LSnPS as extracted from impedance spectroscopy measurements (see Supporting Information for details). **b:** Galvanostatic dc polarization measurement on a symmetric Au|LSnPS|Au cell at 473 K. **c:** Current-voltage curve of the setup Li|liquid electrolyte|Li$_{10}$SnP$_2$S$_{12}$|Au (1mV/s) showing the usable electrochemical window of Li$_{10}$SnP$_2$S$_{12}$ (see Supporting Information S8 for details) **d:** Comparison of the diffusion coefficients obtained from long-range sensitive methods (PFG NMR, impedance spectroscopy) and short-range sensitive methods (NMR relaxometry) for LSnPS and LSiPS. The data derived from NMR relaxometry stem from [a] motional narrowing of the $^7$Li central transition (see Supporting Information S4), [b] motional averaging of the quadrupolar interaction, (see Supporting Information S4), and [c] longitudinal relaxation (see Supporting Information S4). For comparison, the Arrhenius line obtained for LGPS[4] is included as well.

The electrochemical window of Li$_{10}$SnP$_2$S$_{12}$ was examined using cyclic voltammetry on the setup Li|liquid electrolyte|Li$_{10}$SnP$_2$S$_{12}$|Au (see Fig. 5C). This setup avoids direct contact with Li metal and thus possible reactions prior to the measurement. The open circuit voltage (OCV) of such a setup is 2.4 V, which is typical of sulphides. Going from the OCV to lower potentials, a first small reduction peak is observed at 1.4 V (probably the reduction of thiophosphate), the second reduction peak is observed at 0.8 V (reduction of thiostannate), while the first oxidation peak at higher potentials (again starting from the OCV) is observed at 2.6 – 3 V (oxidation of sulphide). A very similar behaviour was observed for Li$_{10}$GeP$_2$S$_{12}$, except that the second reduction peak is observed at 0.6 V (reduction of thiogermanate). The voltage range between 1.4 V and 2.6 V can thus be considered as the range of

thermodynamic stability of these LGPS-type electrolytes (see Supporting Information S8 for further details). However, by applying successively lower voltages starting from the OCV in several cycles, apparently a stable passivation layer (SEI) was formed for $Li_{10}SnP_2S_{12}$ (in the case of $Li_{10}GeP_2S_{12}$ such an SEI readily forms in contact with Li metal at room temperature). After the formation of the SEI, the effective electrochemical window is significantly increased to at least the range between -0.5 V and 4 V as shown in Fig. 5C. The behaviour depicted in Fig. 5C was stable for at least 10 cycles. The SEI is responsible for the large electrochemical window observed for all LGPS-type electrolytes (see Supporting Information S8 for details). The fact that the kinetic stability is sufficient for cell operation is provided by recent work by Kato *et al.*[12] and likewise by the functioning of liquid electrolytes. However, this point needs to be further addressed by future all-solid-state cell engineering.

Fig. 5D summarizes the results obtained from PFG NMR, NMR relaxometry, and conductivity measurements for LSiPS and LSnPS. For the comparison, the jump rates $\tau^{-1}$ and the conductivities $\sigma_{dc}$ were appropriately transformed into diffusion coefficients $D^{uc}$ (uncorrelated diffusion coefficient) and $D^{\sigma}$ (conductivity diffusion coefficient) using the Einstein-Smoluchowski relation $D^{uc}= a^2/6 \times \tau^{-1}$ (jump distance $a$) and the Nernst-Einstein relation $D^{\sigma} = k_B T/(Nq^2) \times \sigma_{dc}$ (number densitiy of Li$^+$ $N$, charge of Li$^+$ $q$, Boltzmann's constant $k_B$). For the calculation, $N$ and an average jump distance of $a \sim 2$Å were deduced from the structure. Obviously, for both $Li_{10}SnP_2S_{12}$ and $Li_{11}Si_2PS_{12}$, the macroscopically observed tracer diffusion can be traced back to 3D Li hopping in the bulk lattice with activation energies of 0.25(1) eV and 0.20(1) eV, respectively. The correlation factor $f$ and the Haven ratio $H_R$ connecting $D^{uc}$ and $D^{\sigma}$ with $D^{tr}$ via $D^{tr} = f \times D^{uc} = H_R \times D^{\sigma}$ are on the order of unity as expected for simple diffusion mechanisms.

**Conclusions**

The Ge-free LGPS-type materials $Li_{11}Si_2PS_{12}$ and $Li_{10}SnP_2S_{12}$ are both ultrafast Li electrolytes and, hence, promising candidates for the development of a new generation of all-solid-state batteries. The Li diffusivity of the Si compound establishes a new record for solid Li conductors, yet the high-pressure treatment and more difficult sinterability make the preparation more costly. In contrast, upscaling the synthesis of the Ge-free ultrafast electrolyte $Li_{10}SnP_2S_{12}$ should be straightforward. Its kinetic stability features appear promising even for the contact with Li. To clarify whether long-time durability of this contact is satisfactory or whether less reactive electrodes need to be used will require long-time all-solid-state cell engineering.

**Notes and references**
[a] Max Planck Institute for Solid State Research, Heisenbergstr. 1, 70569 Stuttgart, Germany.
[b] Department of Chemistry, Ludwig-Maximilans-Universität München, Butenandtstr. 5-13, 81377 München, Germany.
[c] Nanosystems Initiative Munich and Center for Nanoscience, Schellingstr. 4, 81377 München, Germany.

† Electronic Supplementary Information (ESI) available: [experimental, Rietveld refinement results, supplementary tables, NMR relaxometry details, impedance spectroscopy details, determination of electronic transference number, SEM-EDX, electrochemical stability]. See DOI: 10.1039/c000000x/
Financial support was granted by the Max Planck Society, the University of Munich (LMU), the Center for NanoScience (CeNS), and the Deutsche Forschungsgemeinschaft (DFG) through the Cluster of Excellence „Nanosystems Initiative Munich" (NIM). B. V. L. gratefully acknowledges financial support by the Fonds der Chemischen Industrie. We thank K. Schunke and U. Engelhardt for help with the high-pressure experiments.